\documentclass[
    reprint,
    amsmath,
    amssymb,
    aps, 
    superscriptaddress
]{revtex4-1}

\usepackage{float}
\usepackage{bbm}
\usepackage{epsfig}
\usepackage{epstopdf}
\usepackage{graphicx, subfigure}
\usepackage{pgfplots}
\usetikzlibrary{patterns}
\usepgfplotslibrary{groupplots}
\usepackage{amssymb}
\usepackage{amsmath}
\usepackage{scalefnt}
\usepackage{color}
\usepackage[title]{appendix}
\usepackage{hyperref}
\usepackage{qcircuit}
\usepackage{algorithm} 
\usepackage{algpseudocode}
\usepackage[capitalize]{cleveref}
\usepackage{braket,mleftright}
\usepackage{physics}
\usepackage{import}
\usepackage{dcolumn}
\usepackage{csquotes}
\usepackage{tikz}
\usetikzlibrary{calc}
\usepackage{tkz-graph}

\newcommand{\be}{\begin{equation}}
\newcommand{\ee}{\end{equation}}
\newcommand{\bea}{\begin{eqnarray}}
\newcommand{\eea}{\end{eqnarray}}
\def\ket#1{\left| #1 \right\rangle}
\def\bra#1{\left\langle #1 \right|}

\begin{document}

\title{qubit-ADAPT-VQE: An adaptive algorithm for constructing hardware-efficient ans{\"a}tze on a quantum processor}

\author{Ho Lun Tang}
\affiliation{Department of Physics, Virginia Tech, Blacksburg, VA 24061, USA}
\author{V. O. Shkolnikov}
\affiliation{Department of Physics, Virginia Tech, Blacksburg, VA 24061, USA}
\author{George S. Barron}
\affiliation{Department of Physics, Virginia Tech, Blacksburg, VA 24061, USA}
\author{Harper R. Grimsley}
\affiliation{Department of Chemistry, Virginia Tech, Blacksburg, VA 24061, USA}
\author{Nicholas J. Mayhall}
\affiliation{Department of Chemistry, Virginia Tech, Blacksburg, VA 24061, USA}
\author{Edwin Barnes}
\affiliation{Department of Physics, Virginia Tech, Blacksburg, VA 24061, USA}
\author{Sophia E. Economou}
\affiliation{Department of Physics, Virginia Tech, Blacksburg, VA 24061, USA}

\date{\today}

\begin{abstract}
Quantum simulation, one of the most promising applications of a quantum computer, is currently being explored intensely using the variational quantum eigensolver.
The feasibility and performance of this algorithm depend critically on the form of the wavefunction ansatz. Recently in Nat. Commun. 10, 3007 (2019), an algorithm termed ADAPT-VQE was introduced to build system-adapted ans{\"a}tze with substantially fewer variational parameters compared to other approaches. This algorithm relies heavily on a predefined operator pool with which it builds the ansatz. However, Nat. Commun. 10, 3007 (2019) did not provide a prescription for how to select the pool, how many operators it must contain, or whether the resulting ansatz will succeed in converging to the ground state. In addition, the pool used in that work leads to state preparation circuits that are too deep for a practical application on near-term devices.
Here, we address all these key outstanding issues of the algorithm. We present a hardware-efficient variant of ADAPT-VQE that drastically reduces circuit depths using an operator pool that is guaranteed to contain the operators necessary to construct exact ans\"atze. Moreover, we show that the minimal pool size that achieves this scales linearly with the number of qubits.
Through numerical simulations on $\text{H}_4$, LiH and $\text{H}_6$, we show that our algorithm (``qubit-ADAPT") reduces the circuit depth by an order of magnitude while maintaining the same accuracy as the original ADAPT-VQE. A central result of our approach is that the additional measurement overhead of qubit-ADAPT compared to fixed-ansatz variational algorithms scales only linearly with the number of qubits. Our work provides a crucial step forward in running algorithms on near-term quantum devices.
\end{abstract}

\maketitle
\section{Introduction}
Finding the ground state of a many-body interacting electronic Hamiltonian is one of the most important problems in modern quantum chemistry and physics. 
As the dimension of the Hamiltonian scales exponentially with the number of particles, accurate classical simulations can only be performed for systems with few electrons. 
Although many classical computational techniques have been developed to approximate the ground electronic state,
no classical method is available which can perform accurately for arbitrary systems with polynomial effort. 
While density functional theory (DFT) \cite{PhysRev.140.A1133,PhysRev.136.B864} has been tremendously useful for unraveling the microscopic details of weakly correlated molecules and materials, 
the most accurate form (Kohn-Sham DFT \cite{PhysRev.140.A1133}) relies on a single Slater determinant wavefunction, which fails to describe strong correlation with current functionals. 
Unlike DFT, whose accuracy relies on a density functional which is not able to be systematically improved, wavefunction-based methods such as configuration interaction (CI) or coupled-cluster (CC) can be used to describe many-body systems with clear paths to arbitrary accuracy.
However, in the presence of strong correlation such methods incur an exponential computational cost due to the large number of Slater determinants involved. 
Alternatively, methods based on tensor network states \cite{kovyrshinSelfAdaptiveTensorNetwork2017,martiCompletegraphTensorNetwork2010,murgTreeTensorNetwork2015,szalayTensorProductMethods2015,chooFermionicNeuralnetworkStates2019}
(most notably density matrix renormalization group \cite{White1992,Schollwock2005,Chan2011}) can exploit rank sparsity in low-energy states of one-dimensional systems to simulate strongly correlated systems, with polynomial cost. 
This polynomial scaling is lost in higher dimensions, and the computational cost again grows exponentially.
A radically different approach is Feynman's proposal to study quantum systems using quantum computers \cite{Feynman1982}.
Recent reviews provide a comprehensive background and discuss new developments in quantum simulation with quantum computers \cite{mcardle_quantum_2018, doi:10.1021/acs.chemrev.8b00803, bauer_quantum_2020}.

The long-term method for simulating chemical systems with quantum computers is the quantum phase estimation algorithm (PEA) \cite{kitaev_quantum_1995,doi:10.1080/00268976.2011.552441}, which shows an exponential speedup over classical algorithms \cite{PhysRevLett.79.2586,PhysRevLett.83.5162}. 
Since the number of gates, i.e. unitary operations, involved in this algorithm is very large, it requires a long coherent evolution which can only be realized in fault-tolerant quantum computers \cite{Preskill2018quantumcomputingin}. 
These scalable, error-correcting devices may take decades to realize experimentally. 
In the meantime, the community is exploring algorithms that can be applied to existing and near-term processors, namely noisy intermediate-scale quantum (NISQ) devices \cite{Preskill2018quantumcomputingin}. 

A promising algorithm for NISQ hardware is the variational quantum eigensolver (VQE) \cite{Peruzzo2014,McClean_2016}. 
VQE is a hybrid method that combines classical computational power with a quantum processor. 
Based on the variational principle in quantum physics, 
the VQE algorithm constructs a trial wavefunction by applying gates on a quantum device and estimates the average energy by measuring the Hamiltonian on that device. 
This measured energy is then minimized by tuning the quantum circuit.
The circuit parameter optimization is performed by a classical computer, and quantum resources are only used for the classically intractable parts (state preparation and energy evaluation) of the calculation. 
Compared to PEA, VQE has much more modest requirements on the coherence times of the quantum processor, and it has already been realized on NISQ devices, such as superconducting qubits \cite{PhysRevX.8.011021,Kandala2017}, photons \cite{Peruzzo2014}, and trapped ions \cite{PhysRevA.95.020501,PhysRevX.8.031022}. 
The accuracy of VQE is highly dependent on the explicit form of the wavefunction ansatz and can only obtain the exact ground state energy if the ansatz is capable of representing any state in the subspace that contains the ground state. For example, if symmetries of the ground state are known, then one can construct an ansatz that represents any state in the corresponding symmetry subspace \cite{Gard2020}, guaranteeing that the exact ground state can be expressed in terms of the ansatz.

One of the commonly used ans{\"a}tze for VQE is unitary coupled cluster singles and doubles (UCCSD) \cite{RevModPhys.79.291,Yung2014}. 
Stemming from the coupled cluster theory used in chemistry, this unitary version is more suitable for quantum circuit implementation.
The `singles and doubles' in UCCSD means only single and double excitation operators are included in the ansatz, each carrying its own variational parameter.
One drawback of UCCSD is that including all the singles and doubles operators leads to a (potentially unnecessarily) deep circuit and a large number of parameters to optimize.
To reduce this complexity, several alternatives have been proposed which attempt to keep only the most important operators \cite{Grimsley2019,ryabinkinQubitCoupledCluster2018a,ryabinkin_iterative_2019,leeGeneralizedUnitaryCoupled2019,hugginsNonOrthogonalVariationalQuantum2019a,dallaire-demersLowdepthCircuitAnsatz2018c,Romero_2018,190912410Jastrowtype}.
A second drawback is that UCCSD is also generally not exact and suffers from ambiguities in operator ordering upon factorization into a product of exponentiated operators (Trotterization), which is a necessary step in converting the ansatz to a state preparation circuit \cite{Evangelista2019,Grimsley2020}.
Another approach to building the ansatz is to use the most accessible gates in the quantum device, alternating single-qubit gates and two-qubit gates layer by layer.
This is referred to as a hardware-efficient ansatz \cite{Kandala2017}.
Although this approach lessens the demands on the quantum processor, the resulting wavefunction ansatz can lead to difficulties in parameter optimization \cite{McClean2018}.
This problem was addressed by constructing particle-conserving entangling gates instead of ordinary two-qubit gates \cite{PhysRevA.98.022322, PhysRevApplied.11.044092} and symmetry-preserving circuits \cite{Gard2020}. While these fixed-ansatz approaches can be applied to any problem and can reduce the number of variational parameters and circuit depths, further improvement may still be possible by tailoring the ansatz to a given simulation problem.

Recently, a new algorithm that provides a systematic method to build an ansatz dynamically was introduced. 
This algorithm, termed Adaptive Derivative Assembled Pseudo-Trotter (ADAPT) VQE \cite{Grimsley2019}, employs a predetermined pool of operators from which the ansatz is dynamically constructed.
The ansatz is grown iteratively, such that at each step, the operator that affects the energy the most is added to the ansatz. Using fermionic operators as a pool, Ref.~\cite{Grimsley2019} demonstrated that ADAPT-VQE substantially outperforms UCCSD, in terms of both number of variational parameters and accuracy.
This result demonstrates the promise of the ADAPT algorithm.
However, due to the gate overhead of the fermion-to-spin mapping, the operators considered in Ref.~\cite{Grimsley2019} translate to a fairly large number of quantum gates, and, therefore, while the number of parameters is very low, the circuit depth (which is significantly reduced compared to UCCSD) may still be impractically large, limiting the applicability of ADAPT-VQE to NISQ devices. Even more importantly, it is not clear (i) how the operator pool should be chosen in general, (ii) how many operators it should contain, and (iii) what guarantees that the pool is complete, i.e., that it enables convergence to the ground state.

In this paper, we address these issues by introducing a hardware-efficient variant of ADAPT-VQE that substantially reduces both the number of measurements and the circuit depths needed to achieve convergence. 
We term this algorithm qubit-Adaptive Derivative Assembled Problem-Tailored (qubit-ADAPT) VQE, in contrast with the implementation in Ref.~\cite{Grimsley2019}, which we refer to as fermionic-ADAPT in this paper.
Through classical simulations of several different molecules, we demonstrate that compared to fermionic-ADAPT, qubit-ADAPT reduces the circuit depth by an order of magnitude while maintaining the same accuracy. 
Moreover, we introduce a pool completeness criterion that determines whether a given pool will generate an exact ADAPT ansatz. We  {prove} that the minimal number of pool operators that satisfy this condition grows linearly with the number of qubits. This is much smaller than the quartic scaling originally assumed in fermionic-ADAPT, and it  {demonstrates} that the additional measurement overhead of ADAPT-VQE remains modest for larger systems (increasing only linearly over conventional, fixed-ansatz VQEs). Our results pave the way toward both practical and accurate VQE algorithms on NISQ devices.

This paper is organized as follows. First, we briefly review fermionic-ADAPT and estimate the corresponding circuit depth in Sec.~\ref{sec:fADAPT}. In Sec.~\ref{sec:qADAPT}, we introduce qubit-ADAPT and provide a detailed description of the operator pool. In Sec.~\ref{sec:numerics}, we compare the performance of qubit-ADAPT and fermionic-ADAPT through numerical simulations of $\text{H}_4$, LiH and $\text{H}_6$ molecules. We show that the minimal size of the operator pool for qubit-ADAPT scales linearly in the number of qubits in Sec.~\ref{sec:poolreduction}. We conclude in Sec. \ref{sec:conclusion}. Details of the estimate for the circuit depth of fermionic-ADAPT and a constructive proof of the linear scaling of minimal complete pools are included in appendices.

\section{Circuit depth estimate for fermionic-ADAPT}\label{sec:fADAPT}
The ADAPT ansatz is grown by one operator $\hat{\tau}_i=-\hat{\tau}_i^{\dagger}$ at each iteration, and after the $n$-th iteration is given by
\begin{equation}
    \ket{\psi^{ADAPT}(\vec{\theta})}=e^{\theta_k\hat{\tau}_k}\dots e^{\theta_2\hat{\tau}_2}e^{\theta_1\hat{\tau}_1}\ket{\psi^{HF}},
    \label{eq_adapt}
\end{equation}
where $\ket{\psi^{HF}}$ is the HF state.
These operators are selected from an operator pool defined upfront. 
At each iteration, the operator which induces the maximum change to the energy is selected.
This energy response is represented by the gradient of the energy with respect to the corresponding parameter, i.e.
\begin{equation}
    \frac{\partial }{\partial \theta_i}\langle E\rangle=\bra{\psi}[\hat{H},\hat{\tau}_i]\ket{\psi},
    \label{grad}
\end{equation}
which can be measured on the quantum device.
The ansatz keeps growing until the norm of the gradient vector,
\begin{align}
    ||\vec{g}|| = \sqrt{\sum_i \left(\frac{\partial }{\partial \theta_i}\langle E\rangle\right)^2},
\end{align}
is zero or, in practice, smaller than a chosen threshold, $\epsilon$. Compared to ordinary VQE, ADAPT-VQE requires additional measurements to obtain the gradient at each iteration; the number of these measurements is roughly equal to the size of the pool times the number of terms in the Hamiltonian \footnote{Because the gradient depends on the commutator with $\hat{H}$ such that measurements associated with terms in $\hat{H}$ which commute with a specific operator in the pool can be skipped, the actual number of measurements needed is difficult to predict, as it will depend on the operator pool chosen. Getting estimates for this number for various pools will be the focus of future work.}, although the number of measurements needed to compute the mean energy can be decreased using recently developed techniques \cite{190713117Efficient}.

In order to attain a compact circuit for the ADAPT ansatz, we want to minimize both the number of parameters and the circuit depth.
While the first requirement can be satisfied by the general structure of the ADAPT-VQE algorithm, the second one is not guaranteed and depends on the chosen pool.
Using fermionic operators, a \textit{parameter-efficient} operator pool can be constructed from spin-adapted single excitation operators:
\begin{equation}
    \hat{\tau}_1 \propto \ket{\uparrow}_a\bra{\uparrow}_b+\ket{\downarrow}_a\bra{\downarrow}_b-h.c.,
\end{equation}
and double excitation operators:
\begin{equation}
    \begin{split}
        \hat{\tau}_{2,T} \propto & \ket{T,1}_{pq}\bra{T,1}_{rs}+\ket{T,-1}_{pq}\bra{T,-1}_{rs}  \\
    &+ \ket{T,0}_{pq}\bra{T,0}_{rs} - h.c. \\
      \hat{\tau}_{2,S} \propto &\ket{S,0}_{pq}\bra{S,0}_{rs}-h.c.,
      \label{spin_adapt}
    \end{split}
\end{equation}
where $a,b,p,q,r,s$ are spatial orbitals and $T,S$ refer to triplets and singlets formed by $p,q$ or $r,s$.
To implement ADAPT-VQE on qubits, we have to map these fermionic operators to Pauli operators, which has the consequence that a \textit{parameter-efficient} pool is most likely not \textit{gate-efficient}.
In this paper, we employ the Jordan-Wigner (JW) mapping, i.e., $
    \hat{a}_i\rightarrow\prod_{j=0}^{i-1} Z_j (X_i-iY_i).$ Adapting the methods we develop here to other mappings \cite{Bravyi2002,Havlicek2017,Setia2018} will be the subject of future work.

In the JW mapping, a double excitation operator may contain more than one fermionic operator $a^{\dagger}_p a^{\dagger}_q a_r a_s-h.c.$, which is transformed into at most 8 Pauli strings. (A product of four fermionic operators gives 16 terms in total, but each symmetric term is cancelled by its Hermitian conjugate \cite{Romero_2018}.) 
These excitation operators conserve $S^2$, $S_z$ and particle number by summing a number of Pauli strings together, which results in a high gate count per excitation operator.

\begin{figure*}
  \includegraphics[width=17cm]{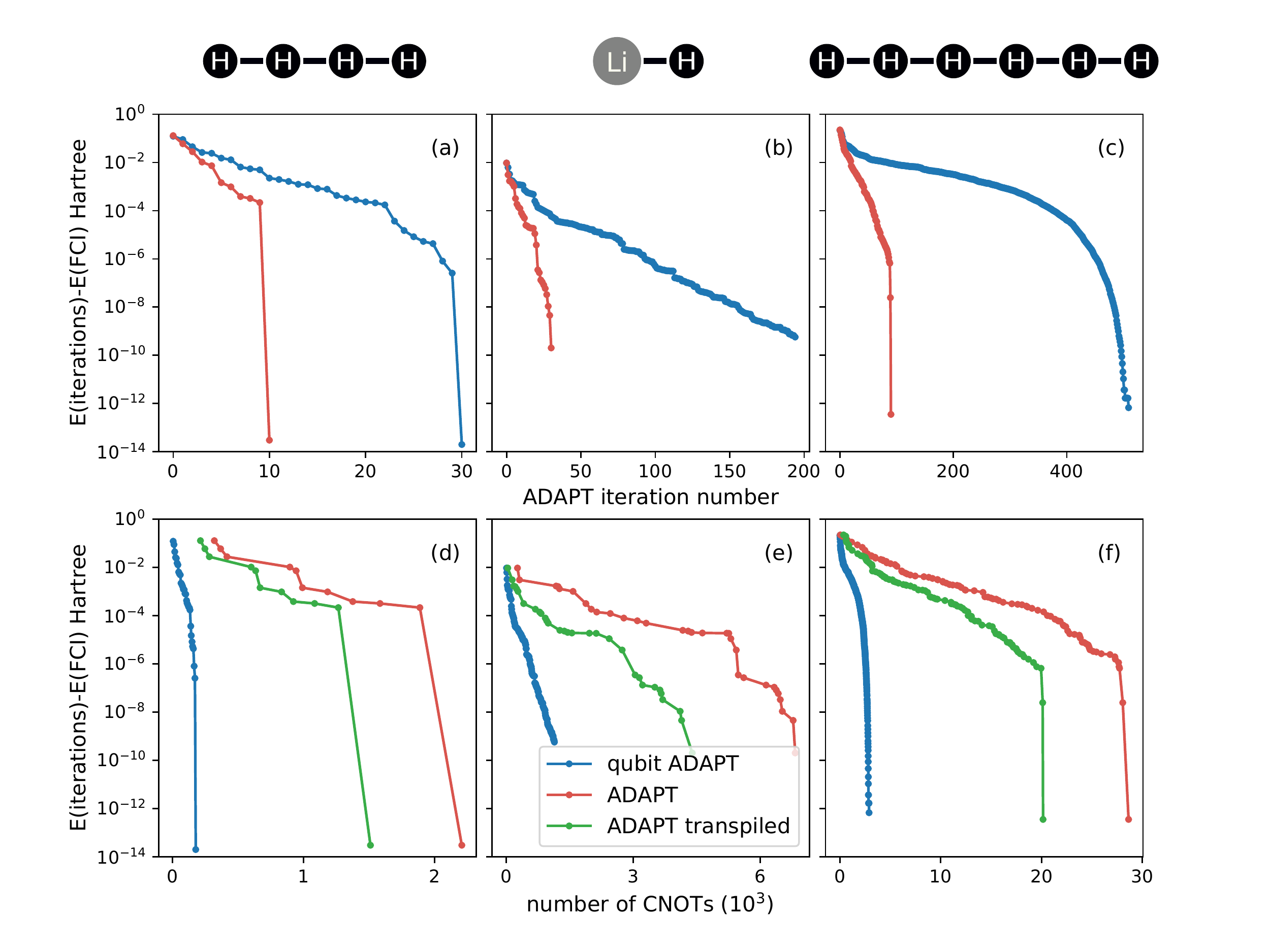}
  \caption{Ground state energy difference between ADAPT and the exact (FCI) result for (a,d) H$_4$ at bond distance $1.5\text{\AA}$, (b,e) LiH at bond distance $2\text{\AA}$, and (c,f) H$_6$ at bond distance $1.5\text{\AA}$. The bond distances are chosen to ensure that correlation effects are substantial.
  All the calculations are performed using an STO-3G basis and start from restricted Hartree-Fock orbitals without using a frozen-core approximation such that we have 8 spin-orbitals for $\text{H}_4$, 12 spin-orbitals for LiH, and 12 spin-orbitals for $\text{H}_6$. Results are shown for qubit-ADAPT (blue), fermionic-ADAPT (orange) and transpiled fermionic-ADAPT (green). Panels (a)-(c) show the energy difference as a function of ADAPT iterations (which is equal to the number of variational parameters), while (d)-(f) show it as a function of the number of CNOTs in the corresponding state preparation circuit. All CNOT gate counts are obtained using the qiskit command \texttt{count\_ops}. The transpiled counts were obtained using \texttt{qiskit.transpile} \cite{Qiskit} with heavy optimization, which includes canceling back-to-back CNOTs and ``commutative cancellation". Although qubit-ADAPT requires more variational parameters, it entails significantly fewer CNOTs compared to fermionic-ADAPT.}
  \label{f_q_compare}
\end{figure*}
        
In order to make a rough estimate of the number of CNOT gates involved in one operator from the fermionic pool,
we consider the generalized singles-doubles excitation fermionic pool.
`Generalized' indicates that the excitations are not restricted to occur only from occupied to virtual orbitals.
Rather,  all combinations of excitations are included. 
To obtain the gate count, we perform first-order Trotterization of each unitary, i.e.,
$ e^{\theta_i \hat{\tau}_i}\approx \prod_j e^{\theta_j \hat{P}_j},$
where $P_j$ are the Pauli strings appearing in $\hat{\tau}_i$ after the JW mapping.
For simplicity, we assume only double excitation operators are picked by the algorithm, which makes this a conservative estimate as any single excitation  
included would result in a smaller CNOT-to-parameter count ratio. 
The number of CNOTs needed for a single Pauli string $\hat{P}_i$ with length $q$ is $2 (q-1)$ \cite{Nielsen:2011:QCQ:1972505}. Here, the length $q$ is the number of non-identity Pauli operators in the string.
The average number of CNOTs involved in a spin-adapted double excitation operator is approximately $    \bar{N}_{Pauli} (6+2 \bar{N}_Z) \bar{N}_{spin}
$, where 
$\bar{N}_{Pauli}$ is the average number of Pauli strings in a doubles operator $a^{\dagger}_p a^{\dagger}_q a_r a_s-h.c.$,
$\bar{N}_Z$ is the average number of Pauli $Z$'s in a Pauli string due to the anticommutation relation of fermionic operators, and
$\bar{N}_{spin}$ is the average number of doubles operators summed in a spin-adapted operator (Eq.\eqref{spin_adapt}).
By Appendix. \ref{ApendA}, these quantities are given by
\begin{align}
    \bar{N}_{Pauli}&=\frac{8m(5m-7)}{m(5m-1)-8}, \\
    \bar{N}_Z&=\frac{12m^3-30m^2+34m-20}{3(5m^2-m-8)}, \\ 
    \bar{N}_{spin}&=\frac{5m^2-m-8}{m^2+m},
\end{align}
where $m$ is the number of spatial orbitals. 
For large $m$, the number of CNOTs in a spin-adapted doubles operator is approximately $64m$.
        
\section{qubit-ADAPT }\label{sec:qADAPT}
Because the spin-adapted fermionic operators each introduce a large number of CNOTs into the state preparation circuit, we are motivated to construct a new pool consisting of operators that involve fewer CNOTs.
One way is to break down the spin-adapted fermionic operators after the JW mapping and choose the individual Pauli strings as the operator pool $
   \hat{\tau}=\hat{P}=i\prod_i p_i, \ \ \   p_i \in \{X,Y,Z\}.
$
This more hardware-efficient choice effectively reduces $\bar{N}_{spin}$ and $\bar{N}_{Pauli}$ to 1, while it contains the same basic elements as the spin-adapted fermionic pool.
An important property of this qubit pool is that it only contains Pauli strings with odd numbers of $Y$'s because the fermionic operators are real, hence $\hat{P}=i\prod_i p_i$ has to be real. We refer to these as ``odd" Pauli strings. The remaining ``even" Pauli strings have no effect on the energy. This is because $\hat H$ is symmetric (time-reversal symmetry is preserved), and so the expectation value of the commutator in Eq. \eqref{grad} will vanish for real $\ket{\psi}$ if $\hat{\tau}_i$ is symmetric (i.e., even). We can therefore restrict $\hat{\tau}_i$ to the odd Pauli strings, which are antisymmetric. Using such operators in the pool also ensures that the ansatz remains real throughout the qubit-ADAPT algorithm, which should be the case when $\hat H$ is time-reversal symmetric.
The length of these strings ranges from 2 to $n$ due to the Pauli $Z$ chain responsible for the fermionic anticommutation relation.
To further reduce gate depth, we remove these Pauli $Z$ chains from the operators.
Numerically, we find that these two pools (with and without $Z$ chains) perform similarly.
The pool without $Z$ chains gives Pauli strings with maximum length $4$. 
The size of this pool is much smaller than the full set of Pauli strings with maximum length $4$, as we only pick operators that already appear in the fermionic pool, which are capable of transforming $\ket{\psi^{HF}}$ to the ground state.
We refer to this reduced pool as the ``qubit pool". Below, we demonstrate that this pool produces significantly shallower state preparation circuits compared to fermionic-ADAPT. We also provide evidence that the size of the qubit pool can be reduced dramatically down to a size that scales only linearly in the number of qubits, which substantially cuts down on the number of measurements needed to run qubit-ADAPT.
        
\subsection{Numerical simulations}\label{sec:numerics}
We compare the performance of the qubit pool to the fermionic pool in terms of the number of parameters and the number of CNOTs for different molecules, $\text{H}_4$, LiH and $\text{H}_6$ (Fig. \ref{f_q_compare}). In each case, we choose a bond distance such that correlation effects are significant: $r=1.5\text{\AA}$ for $\text{H}_4$ and $\text{H}_6$, and $r=2\text{\AA}$ for LiH. All the calculations are performed using an STO-3G basis and start from restricted Hartree-Fock orbitals without using a frozen-core approximation such that we have 8 spin-orbitals for $\text{H}_4$ and 12 spin-orbitals for each of the other two molecules. All CNOT gate counts are obtained using the qiskit command \texttt{count\_ops}, where we consider the case of all-to-all qubit connectivity for concreteness. Although these counts will increase as the connectivity is reduced, we do not expect the relative performance of the two algorithms to change significantly. In the case of fermionic-ADAPT, we show gate counts before and after transpilation is used. The counts before transpilation agree well with the estimates obtained in Sec.~\ref{ApendA}.
        
It is evident from Fig.~\ref{f_q_compare} that in the case of the qubit pool, more parameters are used compared to the fermionic pool. 
On the other hand, the number of CNOTs is reduced significantly, by about an order of magnitude in the case of H$_6$. 
Switching from the fermionic pool to the qubit pool increases the number of parameters in the ansatz, which is the price for compressing the circuit depth. However, the increase in parameter number only increases the required classical computational power during the classical optimization, while the decrease in circuit depth reduces the demands on the quantum processor.
For NISQ devices, the number of CNOTs that can be implemented within the coherence time is very limited, so the ability of qubit-ADAPT to divert more of the computational cost away from the quantum processor and onto the classical optimizer should be advantageous overall.

\begin{figure}
  \includegraphics[width=8.5cm]{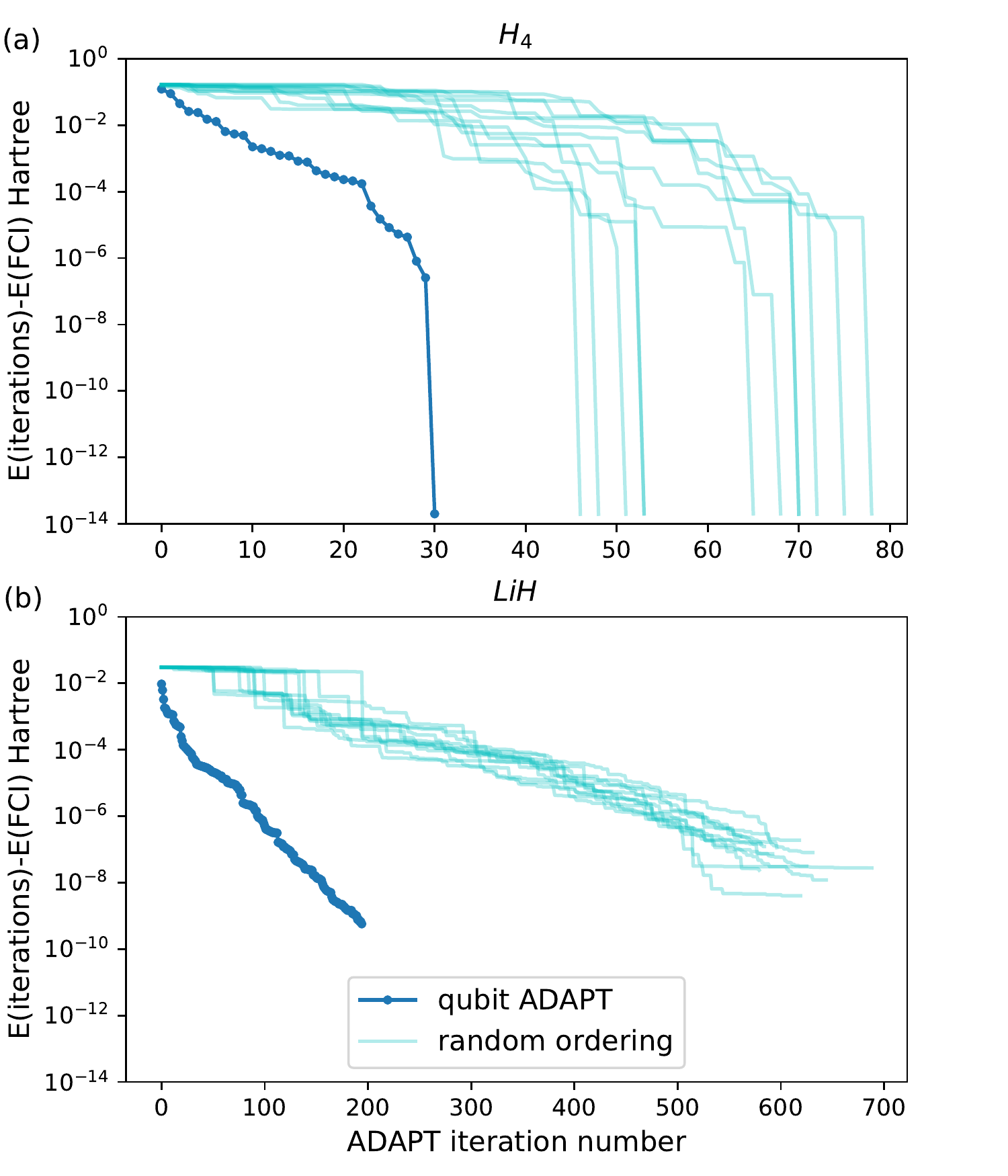}
  \caption{The energy error of qubit-ADAPT versus ans\"atze with random operator orderings as a function of the number of iterations for (a) H$_4$ and (b) LiH. The qubit pool is used in all cases, and the orbital bases and bond distances are chosen as in Fig.~\ref{f_q_compare}.}
  \label{rand}
\end{figure}
        
To evaluate the effectiveness of using the gradient as a means to grow the ADAPT ansatz, we compare the performance of qubit-ADAPT with random operator orderings drawn from the same qubit pool.
In Fig.~\ref{rand}, we see that qubit-ADAPT always converges much faster than the random orderings for both $\text{H}_4$ and LiH.
When random orderings are used for $\text{H}_4$, convergence to the ground state requires 46-78 parameters, compared to only 30 parameters for qubit-ADAPT.
In the case of LiH, the random orderings require more than three times as many parameters as qubit-ADAPT to converge. We can only provide a lower bound on the number of parameters needed to converge the random ordering results in Fig.~\ref{rand}(b) due to the long computational times needed in this case.
These findings suggest that the role of the gradient selection in qubit-ADAPT is crucial for larger problems.

\begin{figure}
  \includegraphics[width=8.5cm]{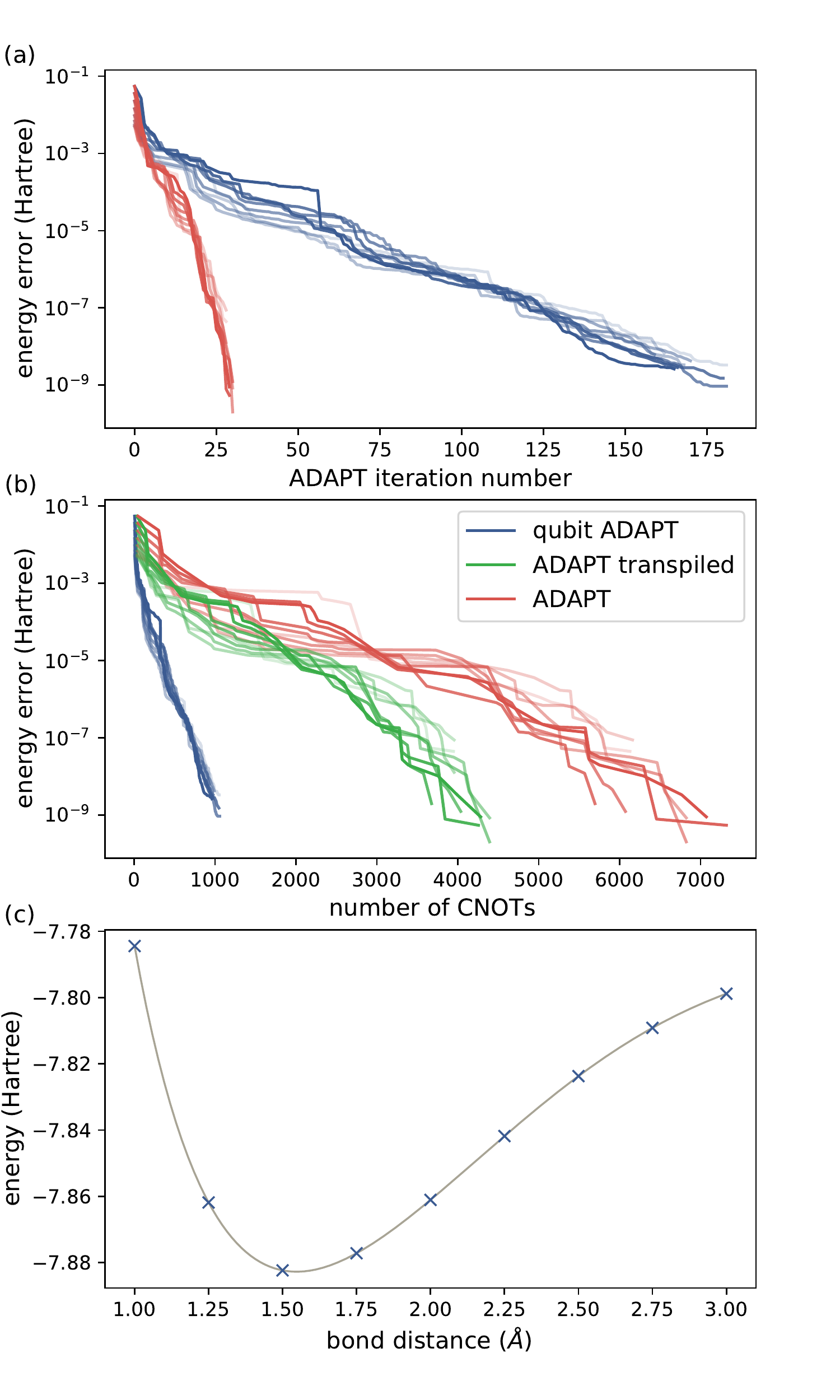}
  \caption{qubit-ADAPT and fermionic-ADAPT energy error of LiH for various bond distances. The bond distance varies from $1{\text{\AA}}$ to $3{\text{\AA}}$ with darker curves in (a), (b) corresponding to larger bond distance. The energy error is plotted against (a) the number of parameters and (b) the number of CNOTs. The performance remains similar across different bond lengths and hence across different amounts of correlation in the ground state. The orbital basis used is the same as in Fig.~\ref{f_q_compare}. (c) The final energies obtained by qubit-ADAPT (points marked by $\times$) for each bond distance considered are plotted along with the exact ground state energies (solid curve).}
  \label{bond distance}
\end{figure}

Fig.~\ref{bond distance} shows that the performance of qubit-ADAPT remains essentially the same across different bond distances. Results are shown for LiH, where the bond distance is varied from $1{\text{\AA}}$ to $3{\text{\AA}}$. The fact that the curves corresponding to different bond lengths largely overlap one another shows that the rate of convergence does not change significantly. Because the amount of correlation in the ground state depends on the bond distance (more correlations tend to arise as the bonds are stretched), this suggests that the convergence of qubit-ADAPT is not sensitive to the strength of correlations. This finding indicates that qubit-ADAPT is a promising approach to studying strongly correlated systems.

\begin{figure}
  \includegraphics[width=8.5cm]{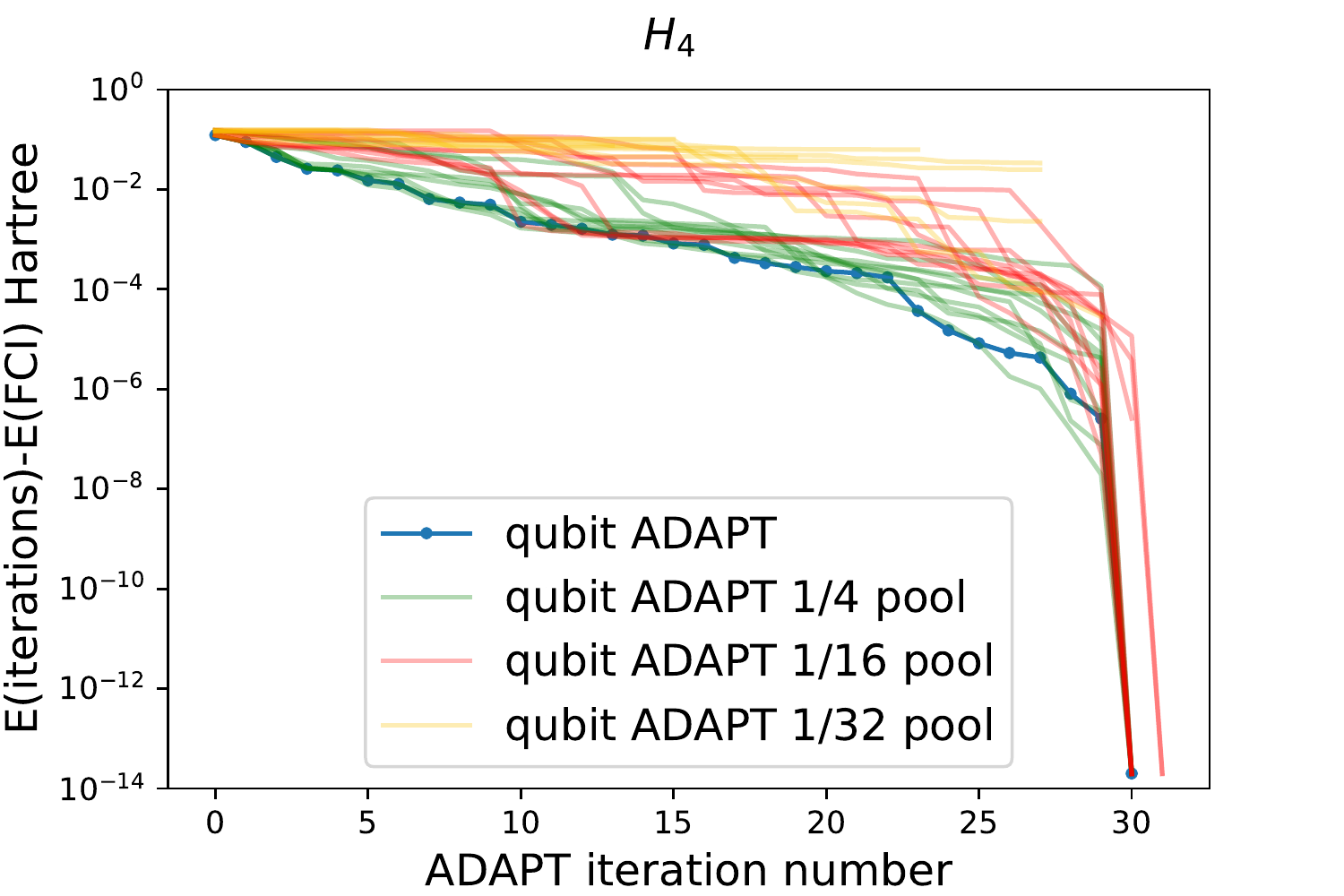}
  \caption{qubit-ADAPT energy error of $\text{H}_4$ using the full qubit pool, and randomly chosen 1/4, 1/16 and 1/32 pools are plotted against the number of iterations. For the 1/4 and 1/16 pools, the algorithm can converge for almost every run. For the 1/32 pool, most of the runs can only reach an energy error of $10^{-3}$. The orbital basis and bond distance are chosen as in Fig.~\ref{f_q_compare}.}
  \label{random pool}
\end{figure}

\subsection{Operator pool reduction}\label{sec:poolreduction}
So far, a drawback of the qubit pool is its large size compared to the fermionic pool, which in turn leads to proportionally more measurements at each iteration. Even though the qubit pool is defined from the fermionic pool, which is a small subset of the full Pauli group, the pool size grows quickly with the number of orbitals.
However, many of these operators are redundant in the sense that eliminating them has no effect on the convergence of the algorithm.
For example, there are pairs of operators that are related by a global rotation, e.g. $X_0Y_1Y_2Y_3$ and $Y_0X_1X_2X_3$, so we only need to keep one of them in the pool; discarding the other does not have any significant effect on the performance of qubit-ADAPT.

The redundancy in the pool can be illustrated by  removing randomly selected operators from the pool and monitoring the impact on convergence. 
First we randomly remove 3/4 of the operators in the pool. As shown in Fig.~\ref{random pool}, despite this large reduction of the pool size, the performance of the algorithm is similar to that with the original pool.
However, as is also evident in Fig.~\ref{random pool}, if we further remove more operators, the pool is sometimes incomplete, and the energy may not converge to the ground state energy.
We can understand the tolerance of the algorithm to the drastic reduction of the pool by studying the Hilbert space spanned by the pool operators.
Starting from the expression for $\ket{\psi^{ADAPT}(\vec{\theta})}$ in Eq. \ref{eq_adapt} and using the Baker-Campbell-Hausdorff formula repeatedly, we obtain
\begin{equation}
    \ket{\psi^{ADAPT}(\vec{\theta})}=e^{\sum_{i}\phi_i A_i}\ket{\psi^{HF}},
    \label{eq_adapt_1}
\end{equation}
where the $A_i$ include all the pool operators and their commutators, and the $\phi_i$ are functions of $\vec\theta$.
Therefore, the Hilbert space spanned by the pool is determined by the set $\{A_i\}$. Note that if the original pool is comprised of odd Pauli strings, then so is $\{A_i\}$.
If the operators in $\{A_i\}$ can transform the reference state to any real state in the $n$-qubit Hilbert space, then the qubit-ADAPT ansatz is guaranteed to be exact, and it is capable of converging to the ground state. (Here, the only symmetry we impose is time-reversal.)
Note that if $\{A_i\}$ includes all the odd Pauli strings (of which there are $2^{n-1}(2^n-1)$, which scales exponentially with the number of qubits), then we could create an arbitrary orthogonal transformation in Eq.~\eqref{eq_adapt_1}. However, spanning the Hilbert space requires only a subset of the odd Pauli strings, because only $2^n-1$ real parameters are required to create an arbitrary real state. In particular, we need $\{A_i\}$ to be such that for an arbitrary state $\ket{\psi}$, the states $A_i\ket{\psi}$ form a complete basis. In this case, we refer to $\{A_i\}$ as a complete basis of operators.

\begin{figure}
  \includegraphics[width=7cm]{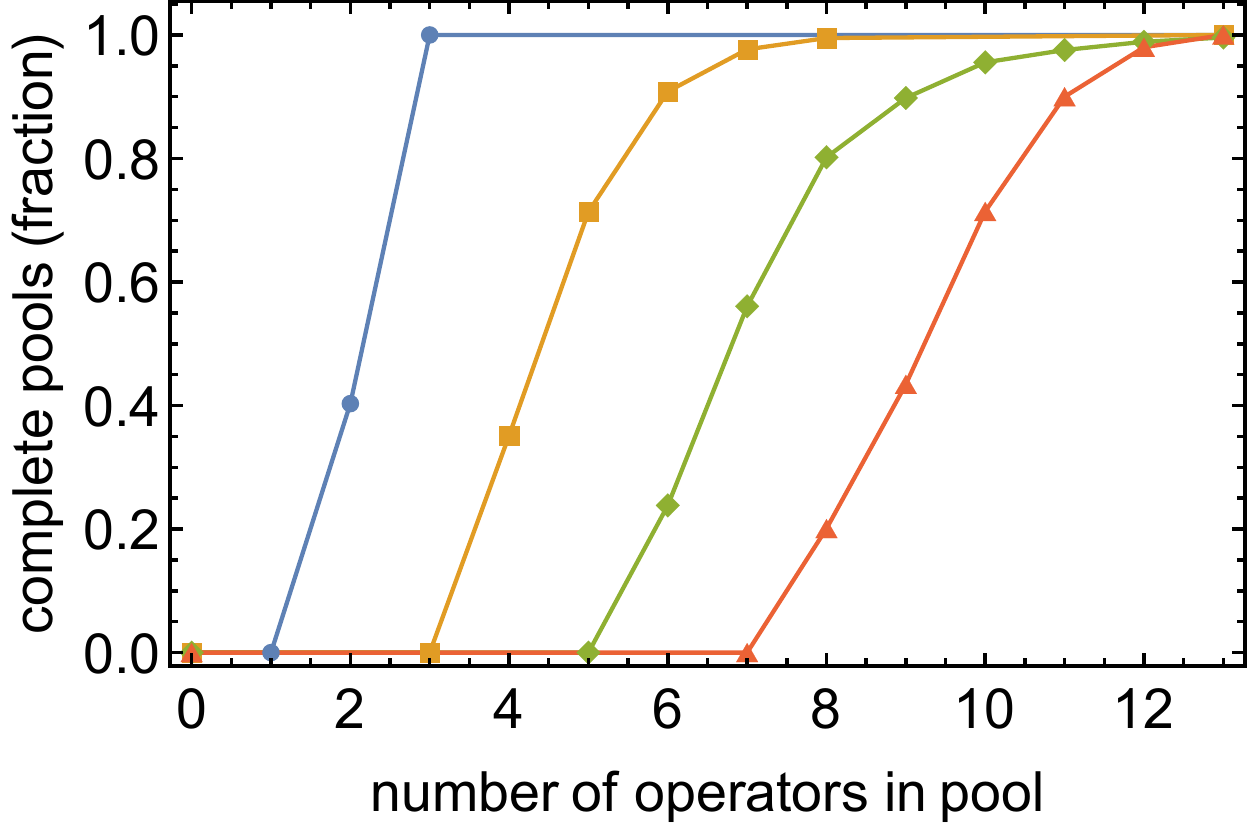}
  \caption{The fraction of pools that are complete as a function of pool size for two (blue circles), three (yellow squares), four (green diamonds), and five (red triangles) qubits.}
  \label{complete_pools}
\end{figure}

The problem then is to determine the minimal pools that produce a complete basis of operators. For a given pool, we define the overlap matrix as $M_{ij}=\bra{\psi}A_i^{\dagger}A_j\ket{\psi}$ where $\ket{\psi}$ is an arbitrary real state.
If the rank of $M$ satisfies $r(M)\ge 2^n-1$, this pool is called complete. To determine the smallest complete pools, we randomly generate many different pools of increasing size and compute $r(M)$ in each case to test for completeness. We did this for up to 7 qubits. Our numerical investigations reveal that, surprisingly, the minimal pool size required for the overlap matrix to have the required rank of $2^n-1$ is only $2n-2$. This is evident in Fig.~\ref{complete_pools}, which shows the fraction of pools that are complete for various pool sizes and numbers of qubits. In each case, complete pools are found for pool sizes that contain at least $2n-2$ operators. Note that not only is this much smaller than the Hilbert space, it is also much smaller than the size of the fermionic pool, which scales like $n^4$. 

Randomly selecting pools of size $2n-2$ and testing for completeness by computing $r(M)$ is a numerically intensive process that quickly becomes infeasible as the number of qubits increases. This is further exacerbated by the fact that the fraction of complete pools of size $2n-2$ becomes smaller as $n$ increases, as is evident in Fig.~\ref{complete_pools}, which raises the question of whether complete pools of size $2n-2$ even exist for large values of $n$. In Appendix~\ref{app:proof}, we prove analytically using induction that complete pools of size $2n-2$ exist for any $n$. In fact, the proof is constructive: We present two families of minimal pools that are provably complete for any number of qubits. One of these pools, which we call $\{V_j\}_n$ where $j=1,\ldots,2n-2$, is defined recursively as $\{V_j\}_n=\{Z_n\{V_k\}_{n-1},iY_n,iY_{n-1}\}$, starting from the pool for $n=2$ qubits: $\{V_j\}_2=\{iZ_2Y_1,iY_2\}$. This pool is comprised of generators for single-qubit $Y$ rotations and conditional $Y$ rotations, and it contains operators that act on up to all $n$ qubits. In Appendix~\ref{app:proof}, we prove that these operators are sufficient to rotate any real state to any other real state in the Hilbert space. As shown in Appendix~\ref{app:mappools}, $\{V_j\}_n$ can be mapped to a second family of minimal complete pools, which we call $\{G_j\}_n$, that has a very local structure. This pool contains all two-qubit operators of the form $iZ_{k+1}Y_k$ that act on two neighboring qubits labeled by $k$ and $k+1$. There are $n-1$ such operators. $\{G_j\}_n$ also contains all single-qubit Pauli $iY$ operators except on the first qubit. Therefore, this pool contains $2n-2$ operators. In Appendix~\ref{app:mappools}, we show that the $G_j$ can be obtained from commutators of the $V_j$ for any $n$, and so the completeness of the former follows from that of the latter. 

\begin{figure}
  \includegraphics[width=8.5cm]{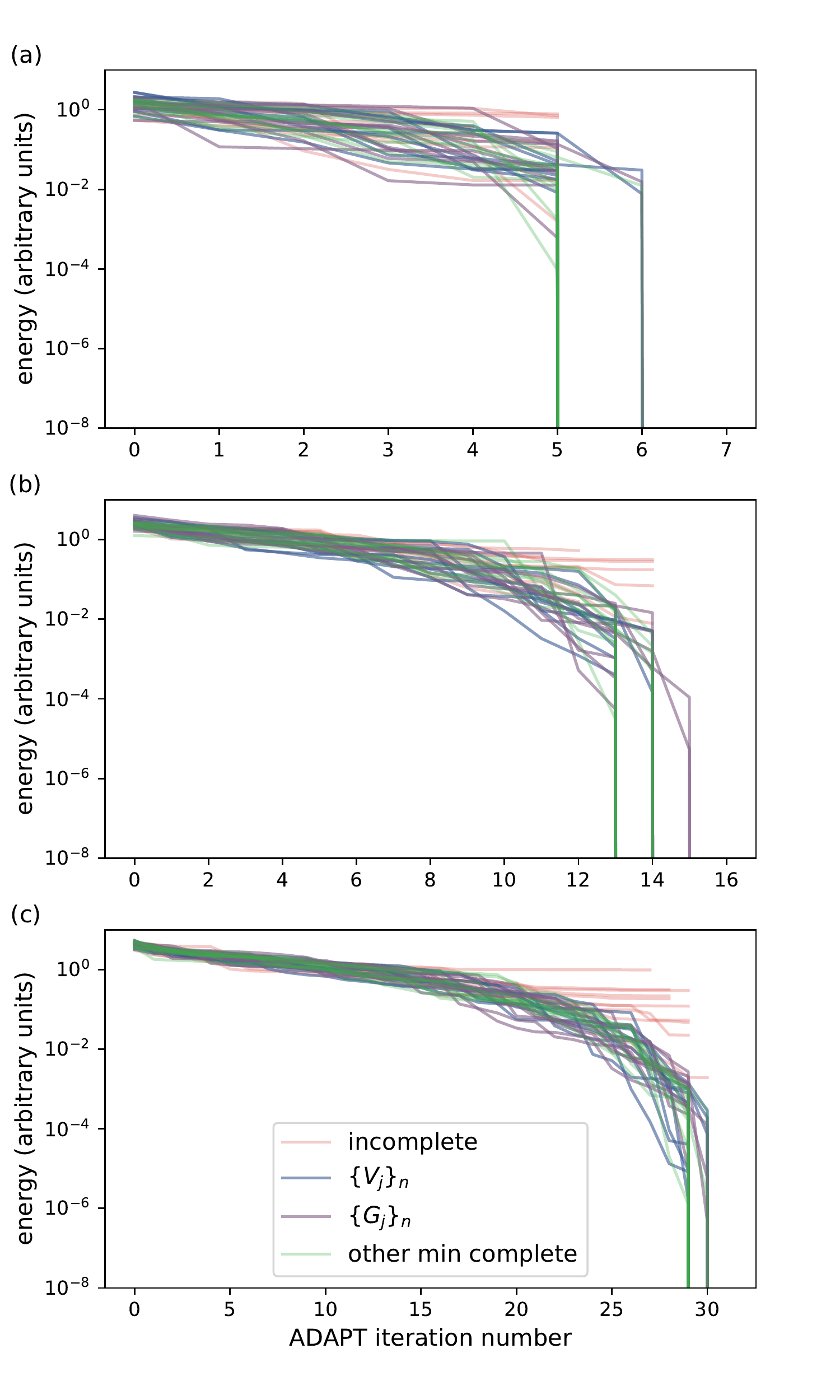}
  \caption{Energy error curves for three different kinds of minimal complete pools versus incomplete pools. Results for 3, 4 and 5-qubit random Hamiltonians are shown in panels (a), (b) and (c) respectively. }
  \label{complete}
\end{figure}

To investigate the importance of the pool being complete, we run qubit-ADAPT for random real Hamiltonians of 3, 4 and 5 qubits with random initial states and for pools consisting of $2n-2$ operators randomly chosen from the set of odd Pauli strings. We also run simulations using the minimal complete pools $\{V_j\}_n$ and $\{G_j\}_n$. The operator coefficients in each Hamiltonian (which is taken to be real and symmetric) are obtained by sampling uniformly in the range $[-2,2]$ with 10 samples for each of these 3 cases. The corresponding energy error curves are illustrated in Fig.~\ref{complete}. Each curve is the result for a different random Hamiltonian. All of the curves that fail to converge correspond to incomplete pools. For these cases, even though the gradient goes to zero the ground state is not reached because important operators are never generated. On the other hand, the runs with complete pools always converge, highlighting the importance of this criterion. For the cases considered, we find that 20-40\% of pools containing $2n-2$ operators are complete. These findings shed light on the question of what constitutes a good operator pool for ADAPT-VQE. These points were not appreciated in the original paper \cite{Grimsley2019}, and moreover, they suggest that the fermionic pool used in that work is overcomplete. Furthermore, the fact that the minimal pool size is linearly proportional to $n$ means that the number of additional measurements needed for each step of qubit-ADAPT is also linear in $n$. Thus, the extra measurement overhead of qubit-ADAPT remains modest as the problem size increases.

\section{Conclusions}\label{sec:conclusion}
In conclusion, we introduced a more efficient and NISQ-compatible version of the ADAPT-VQE algorithm, called qubit-ADAPT. The basic idea of qubit-ADAPT is to use a pool consisting of Pauli strings (rather than fermionic operators) so that the number of CNOT gates associated with each pool operator is reduced. We established a completeness condition that guarantees that a pool will generate an exact ADAPT ansatz, and we {proved} that the smallest possible pool that obeys this criterion scales linearly with the number of qubits. {We also provide a constructive approach to generating minimal complete pools for an arbitrary number of qubits.} These results remove the ad-hoc elements of the ADAPT algorithm and lead to a substantial reduction in the depths of state preparation circuits and in the number of measurements needed to run ADAPT-VQE on realistic hardware. 

\section*{Acknowledgments}
This research was supported by the National Science Foundation (Award No. 1839136) and the US Department of Energy (Award No. DE-SC0019199).

\appendix

\section{CNOT estimation}\label{ApendA}

From the structure of the pool operators, we can try estimating the number of CNOT gates as a function of the number of variational parameters. These estimates are described below.

\subsection*{Qubit ADAPT}
In order to get a conservative estimate on the CNOT count, we assume that all operators picked are four-qubit pauli strings, since exponentiating an n-qubit Pauli string requires $2(n-1)$ CNOT gates, the number of CNOT gates is
\begin{equation}
    \bar{N}_{CNOT}=6
\end{equation}

\subsection*{Fermionic ADAPT}
In the case of fermionic-ADAPT, the number of CNOTs used can be estimated as
\begin{equation}
    \bar{N}_{CNOT}= (6+2\times \bar{N}_{Z})\times \bar{N}_{Pauli} \times \bar{N}_{spin}
\end{equation}
where $\bar{N}_Z$ is the average number of Pauli-Z used for fixing the antimsymmetry of fermionic operators, $\bar{N}_{spin}$ is the average number of fermionic operators in each spin-adapted group and $\bar{N}_{Pauli}$ is the average number of Pauli strings from each fermionic operators. 

\begin{equation}
    \begin{split}
        \bar{N}_{spin}&=\frac{\text{Total \# of fermionic terms ($a^{\dagger}_{p}a^{\dagger}_{q}a_ra_s$)}}{\text{Total spin adapted double ops in pool}}\\
        &:\text{\# of fermionic terms in a spin adapted op} \\
        \bar{N}_{Pauli}&=\frac{\text{Total number of Pauli strings after JW}}{\text{Total fermionic terms}}\\
        &:\text{\# of Pauli strings from each fermionic term} \\ 
        \bar{N}_{Z}&=\frac{\text{Total length of Pauli Z from all fermionic}}{\text{Total \# of fermionic terms}} \\
        &:\text{increases the length of Pauli strings}
    \end{split}
    \label{average}
\end{equation}

To compute these total counts, we can first classify the combination of spatial orbitals in 5 groups:
\begin{enumerate}
    \item $p\neq q\neq r\neq s$
    \item One of $p,q$ = One of $r,s$
    \item $p =q \neq r \neq s$ 
    \item $p =q = r \neq s$
    \item $p=q \neq r=s$.
\end{enumerate}
Since some of them can have both triplet and singlet while the others can only have singlet.
And the number of fermionic terms in spin adapted grouping also depends on the combination of spatial orbitals.

For group 1, the triplet operator reads as
\begin{align*}
\hat{\tau}_{2,T} \propto & \ket{T,1}_{pq}\bra{T,1}_{rs}+\ket{T,-1}_{pq}\bra{T,-1}_{rs} \\
&+ \ket{T,0}_{pq}\bra{T,0}_{rs} - h.c. \\
= & \ket{pq}\bra{rs} + \frac{1}{2}\left(\ket{p\Bar{q}}+\ket{\Bar{p}q}\right)\left(\bra{r\Bar{s}}+\bra{\Bar{r}s}\right) \\
&+ \ket{\Bar{p}\Bar{q}}\bra{\Bar{r}\Bar{s}} - h.c. \\
\rightarrow & a_{p}^\dagger a_{q}^\dagger a_{r} a_{s} + \frac{1}{2}\Big(a_{p}^\dagger a_{\Bar{q}}^\dagger a_{r} a_{\Bar{s}} + a_{p}^\dagger a_{\Bar{q}}^\dagger a_{\Bar{r}} a_{s} \\
&+ a_{\Bar{p}}^\dagger a_{q}^\dagger a_{r} a_{\Bar{s}} + a_{\Bar{p}}^\dagger a_{q}^\dagger a_{\Bar{r}} a_{s}\Big) + a_{\Bar{p}}^\dagger a_{\Bar{q}}^\dagger a_{\Bar{r}} a_{\Bar{s}} - h.c.,
\end{align*}
so there are 6 terms in total (12 terms including Hermitian conjugate). For the singlet, 
\begin{align*}
\hat{\tau}_{2,S} \propto &\ket{S,0}_{pq}\bra{S,0}_{rs}-h.c. \\
=&\frac{1}{2}\left(\ket{p\Bar{q}}-\ket{\Bar{p}q}\right)\left(\bra{r\Bar{s}}-\bra{\Bar{r}s}\right)-h.c. \\
\rightarrow&\frac{1}{2}\Big(a_{p}^\dagger a_{\Bar{q}}^\dagger a_{r} a_{\Bar{s}} - a_{p}^\dagger a_{\Bar{q}}^\dagger a_{\Bar{r}} a_{s} \\
&- a_{\Bar{p}}^\dagger a_{q}^\dagger a_{r} a_{\Bar{s}} + a_{\Bar{p}}^\dagger a_{q}^\dagger a_{\Bar{r}} a_{s}\Big)-h.c.,
\end{align*}
so there are 4 terms in total (8 terms including Hermitian conjugate).
For group 2, the triplet operator reads as (for $q = r$)
\begin{align*}
\hat{\tau}_{2,T}\rightarrow & a_{p}^\dagger a_{q}^\dagger a_{q} a_{s} + \frac{1}{2}\Big(a_{p}^\dagger a_{\Bar{q}}^\dagger a_{q} a_{\Bar{s}} + a_{p}^\dagger a_{\Bar{q}}^\dagger a_{\Bar{q}} a_{s} \\
&+ a_{\Bar{p}}^\dagger a_{q}^\dagger a_{q} a_{\Bar{s}} + a_{\Bar{p}}^\dagger a_{q}^\dagger a_{\Bar{q}} a_{s}\Big) + a_{\Bar{p}}^\dagger a_{\Bar{q}}^\dagger a_{\Bar{q}} a_{\Bar{s}} - h.c.,
\end{align*}
while the singlet operator is
\begin{align*}
\hat{\tau}_{2,S}\rightarrow &\frac{1}{2}\Big(a_{p}^\dagger a_{\Bar{q}}^\dagger a_{q} a_{\Bar{s}} - a_{p}^\dagger a_{\Bar{q}}^\dagger a_{\Bar{q}} a_{s} \\
&- a_{\Bar{p}}^\dagger a_{q}^\dagger a_{q} a_{\Bar{s}} + a_{\Bar{p}}^\dagger a_{q}^\dagger a_{\Bar{q}} a_{s}\Big)-h.c..
\end{align*}
For group 3, there is only singlet
\begin{align*}
\hat{\tau}_{2,S}\rightarrow &\frac{1}{2}\Big(a_{p}^\dagger a_{\Bar{p}}^\dagger a_{r} a_{\Bar{s}} - a_{p}^\dagger a_{\Bar{p}}^\dagger a_{\Bar{r}} a_{s} \\
&- a_{\Bar{p}}^\dagger a_{p}^\dagger a_{r} a_{\Bar{s}} + a_{\Bar{p}}^\dagger a_{p}^\dagger a_{\Bar{r}} a_{s}\Big)-h.c. \\
=&a_{p}^\dagger a_{\Bar{p}}^\dagger a_{r} a_{\Bar{s}} + a_{p}^\dagger a_{\Bar{p}}^\dagger a_{\bar{r}} a_{s} - h.c..
\end{align*}
For group 4, there is only singlet
\begin{align*}
\hat{\tau}_{2,S}\rightarrow &\frac{1}{2}\Big(a_{p}^\dagger a_{\Bar{p}}^\dagger a_{p} a_{\Bar{s}} - a_{p}^\dagger a_{\Bar{p}}^\dagger a_{\Bar{p}} a_{s} \\
&- a_{\Bar{p}}^\dagger a_{p}^\dagger a_{p} a_{\Bar{s}} + a_{\Bar{p}}^\dagger a_{p}^\dagger a_{\Bar{p}} a_{s}\Big)-h.c. \\
=&a_{\bar{p}}^\dagger a_{p}^\dagger a_{p} a_{\Bar{s}} + a_{p}^\dagger a_{\Bar{p}}^\dagger a_{\bar{p}} a_{s} - h.c..
\end{align*}
For group 5, there is only singlet
\begin{align*}
\hat{\tau}_{2,S}&\rightarrow \frac{1}{2}\Big(a_{p}^\dagger a_{\Bar{p}}^\dagger a_{r} a_{\Bar{r}} - a_{p}^\dagger a_{\Bar{p}}^\dagger a_{\Bar{r}} a_{r} \\
&- a_{\Bar{p}}^\dagger a_{p}^\dagger a_{r} a_{\Bar{r}} + a_{\Bar{p}}^\dagger a_{p}^\dagger a_{\Bar{r}} a_{r}\Big)-h.c. \\
=&2a_{p}^\dagger a_{\Bar{p}}^\dagger a_{r} a_{\Bar{r}} - h.c..
\end{align*}
Then we can perform Jordan-Wigner transformation to each of these terms. Each anti-Hermitian pair of fermionic operators having 4 different indices, i.e. $a_{i}^{\dagger}a_{j}^{\dagger}a_{k}a_{l}-h.c.$, would result in 8 Pauli strings, i.e.
\begin{align*}
a_{i}^{\dagger}a_{j}^{\dagger}a_{k}a_{l}-h.c. = 2i\prod_n Z_n &\Big(Y_iX_jX_kX_l+X_iY_jX_kX_l\\
&-X_iX_jY_kX_l-X_iX_jX_kY_l \\
    &-X_iY_jY_kY_l-Y_iX_jY_kY_l\\
    &+Y_iY_jX_kY_l+Y_iY_jY_kX_l\Big).
\end{align*}
The other case is 2 of the 4 indices are the same, 
\begin{align*}
a_{i}^{\dagger}a_{j}^{\dagger}a_{j}a_{l}-h.c.&= \prod_n Z_n (X_i+iY_i)Z_j(X_l-iY_l) -h.c.\\
&= 4i\prod_n Z_n \Big(Y_iZ_jX_l-X_iZ_jY_l\Big),
\end{align*}
only 2 Pauli strings left.
From the expression of the 5 cases' fermionic operators, we can then determine the average number of Pauli strings per anti-Hermitian pair for each case.

\begin{center}
\begin{tabular}{| c | c | c | c | c | c |}
\hline
    Group      & 1  & 2 & 3  & 4 & 5 \\ 
 \hline
 \# of fermi ops (T) & 6 & 6 & NA  & NA & NA \\
 \hline
 \# of fermi ops (S) & 4  & 4 & 2 & 2  & 1\\  
 \hline
Average \# of Paulis (T) & 8   & (4$\times$2+2$\times$8)/6=4 & NA & NA  & NA \\  
 \hline
Average \# of Paulis (S)  &  8   & (2$\times$2+2$\times$8)/4=5 & 8 &  2   & 8 \\  
\hline
\end{tabular}
\end{center}
\begin{table*}
\begin{tabular}{| c | c | c | c | c | c |}
\hline
  Group    & 1  & 2 & 3\\ 
 \hline
 \# of combinations & $\binom{m}{2}\binom{m-2}{2}/2$ & $m\binom{m-1}{2}$ & $m\binom{m-1}{2}$ \\
 \hline
 \# of spin groups    & $\binom{m}{2}\binom{m-2}{2}/2\times 2$    & $m\binom{m-1}{2}\times2$ & $m\binom{m-1}{2}$\\  
 \hline
 \# of fermi ops  & $\binom{m}{2}\binom{m-2}{2}/2\times (6+4)$   & $m\binom{m-1}{2}\times(6+4)$ &  $m\binom{m-1}{2}\times 2$ \\  
 \hline
 \# of Pauli strings  &  $\binom{m}{2}\binom{m-2}{2}/2\times (6+4)\times 8$   & $m\binom{m-1}{2}\times(6\times 4+4\times 5)$ & $m\binom{m-1}{2}\times 2\times 8$ \\  
 \hline
 \# of Pauli Z  & $\frac{m}{2}(2m^4-15m^3+40m^2-45m+18)$ & $\frac{10m}{3}(m^3-4m^2+5m-2)$ & $\frac{2m}{3}(m^3-4m^2+5m-2)$ \\
 \hline
\end{tabular}
\end{table*}
\begin{table*}
\begin{tabular}{| c | c | c | c | c | c |}
\hline
  Group     & 4 & 5 & Total \\ 
 \hline
 \# of combinations &  $2\binom{m}{2}$ & $\binom{m}{2}$ & $m(m^3+2m^2-m-2)/8$ \\
 \hline
 \# of spin groups    & $2\binom{m}{2}$ & $\binom{m}{2}$ & $m^2(m^2-1)/4$\\  
 \hline
 \# of fermi ops  &  $2\binom{m}{2}\times2$ & $\binom{m}{2}$ & $m(5m^3-6m^2-7m+8)/4$ \\  
 \hline
 \# of Pauli strings  & $2\binom{m}{2}\times2\times2$ & $\binom{m}{2}\times8$ & $2m(5m^3-15m^2+14m-4)$ \\  
 \hline
 \# of Pauli Z  &  $\frac{2m}{3}(2m^2-3m+1)$ & 0 & $m(6m^4-21m^3+32m^2-27m+10)/6$ \\
 \hline
\end{tabular}
\caption{For each group, the number of combinations are calculated. For each combination, it has 2 spin group if it has both tripplet and singlet operators, only 1 spin group if it only has singlet. For each singlet and triplet operator, the number of fermionic terms are calculated. And then we calculate the number of Paulis in each term. Finally, we counted the length of Pauli Z.}
\label{tab}
\end{table*}
For the length of Pauli Z, it is equivalent to count the interval between $i$ and $j$ and between $k$ and $l$ where $i\leq j\leq k\leq l$, we calculate them for separate cases, for $m$ spatial orbitals, 
\begin{enumerate}
  \item $p\neq q\neq r\neq s$:
  \\\\The total number of Pauli-Z is given by the sum
  \begin{align*}
      \frac{1}{2}\binom{4}{2}\times (6+4)\times\sum_{m_1<m_2<m_3<m_4} \Big[&(2m_4-2m_3-1)
      \\&+(2m_2-2m_1-1)\Big] \\
      =\frac{m}{2}(2m^4-15m^3+40m^2-45m+&18).
  \end{align*}
  The prefactor $\frac{1}{2}\binom{4}{2}$ is counting ways we choose 2 of $\{p,q,r,s\}$ to be in dagger side, while the prefactor $(6+4)$ is from the number of fermionic operators for each combination.
  If we have $\{i,j,k,l\}=\{1,2,3,4\}$, which is in group 5 , the Jordan-Wigner transformation would bring a Pauli Z chain $Z_1Z_2Z_3$ for the creation/annihilation operator on index $4$ and a Pauli Z chain $Z_1Z_2$ for operator on index $3$, so only $Z_3$ can survive, but it would combine with the $X_3\pm iY_3$ from the operator on index $3$, so there is no Pauli Zs effectively.
  Therefore, we should count the number of integers can be fitted in the intervals $2m_4-2m_3$ and $2m_2-2m_1$, that explains the terms $-1$ in the calculation.
  \\\\
  In this counting, we treat all the spin orbitals to be spin-up (even number indices) even though each term in $\ket{T,0}\bra{T,0}$ and $\ket{S,0}\bra{S,0}$ have 2 spin-down and 2 spin-up operators. It is because whenever there are terms with both spin-up and spin-down spin orbitals, there would be another term with the spin totally flipped, so the counting of these two terms would cancel each other.
  For example, consider $p<q<r<s$, the $\ket{T,0}\bra{T,0}$ terms read as $\frac{1}{2}\left(a_{p}^\dagger a_{\Bar{q}}^\dagger a_{r} a_{\Bar{s}} + a_{p}^\dagger a_{\Bar{q}}^\dagger a_{\Bar{r}} a_{s} + a_{\Bar{p}}^\dagger a_{q}^\dagger a_{r} a_{\Bar{s}} + a_{\Bar{p}}^\dagger a_{q}^\dagger a_{\Bar{r}} a_{s}\right)$ - h.c., the first term and the last term are different by a total spin flip, the number of Pauli Z would be $2s+1-2r-1+2q+1-2p-1$ and $2s-(2r+1)-1+2q-(2p+1)-1$, so the $+1$ due to the odd index switches sign by the spin flip.
  Therefore, the average number of Pauli Z of this pair is the $2s-2r-1+2q-2p-1$, which is the number of Pauli Z in the all spin-up term $a_{p}^\dagger a_{q}^\dagger a_{r} a_{s}$ (or the all spin-down term).
  \item One of $p,q$ = One of $r,s$
  \\\\The total number of Pauli-Z is given by the sum
  \begin{align*}
      &10\times\Big[\sum_{m_1=m_2<m_3<m_4}(2m_4-2m_3-1) \\
      &+ \sum_{m_1<m_2=m_3<m_4}(2m_4-2m_3-1+2m_3-2m_1-1) \\
      &+\sum_{m_1<m_2<m_3=m_4}(2m_2-2m_1-1)\Big] \\
      =&\frac{10m}{3}(m^3-4m^2+5m-2).
  \end{align*}
  Here, we don't have the counting prefactor in group 1 as we already decided that one of the two identical indices is in the dagger side.
  \item $p =q \neq r \neq s$
  \\\\The total number of Pauli-Z is given by the sum
  \begin{align*}
      &2\times\Big[\sum_{m_1=m_2<m_3<m_4}(2m_4-2m_3-1) \\
      &+ \sum_{m_1<m_2=m_3<m_4}(2m_4-2m_3-1+2m_3-2m_1-1)  \\
      &+\sum_{m_1<m_2<m_3=m_4}(2m_2-2m_1-1)\Big] \\
      =&\frac{2m}{3}(m^3-4m^2+5m-2).
  \end{align*}
  It is only different from case 2 by the prefactor $2$ instead of $10$ as it has 2 fermionic operators in total.
  \item $p =q = r \neq s$
  \\\\The total number of Pauli-Z is given by the sum
  \begin{align*}
      &2\times\Big[\sum_{m_1=m_2=m_3<m_4}(2m_4-2m_1-1) \\
      &+ \sum_{m_1<m_2=m_3=m_4}(2m_4-2m_1-1) \Big] \\
      =&\frac{2m}{3}(2m^2-3m+1).
  \end{align*}
  \item $p=q \neq r=s$.
  \\\\  Since the operators have a form $a^{\dagger}_{p}a^{\dagger}_{\Bar{p}}a_{r}a_{\Bar{r}}$, they don't have Pauli Z.
\end{enumerate}
Using Eq.\eqref{average} and Table \ref{tab}, the average number can then be calculated as follow
\begin{align}
    &\bar{N}_{spin}=\frac{5m^2-m-8}{m^2+m} \\
    &\bar{N}_{Pauli}=\frac{8(5m^3-15m^2+14m-4)}{5m^3-6m^2-7m+8} \\
    &\bar{N}_{Z}=\frac{12m^3-30m^2+34m-20}{3(8+m-5m^2)}.
\end{align}
We can estimates the CNOTs required for H4 molecule (m=4) is 
\begin{align*}
    N_{CNOT}&\approx N_{pars}\times N_{Pauli}\times (6+2\times N_Z)\times N_{spin} \\
    &=11\times N_{Pauli}\times(6+2\times N_Z)\times N_{spin} \\
    &=1928.41
\end{align*}
which is very close to the number we obtained.

\begin{center}
\begin{tabular}{|c|c|c|c|}
     & H4 & LiH & H6 \\
     \hline
    estimated \# of CNOT per param  & 175 & 303 & 303 \\
    \hline
    \# of param  & 11 & 30 & 91 \\
    \hline
    total estimated \# of CNOT & 1928 & 9078 & 27536 \\
    \hline
    \# of CNOT  & 2208 & 6824 & 28632
\end{tabular}
\end{center}
This estimation is accurate when all the spin orbitals are evenly explored (H6 and H4), i.e. no `core' orbitals.

\section{Constructive proof of minimal complete pools}\label{app:proof}

In this appendix, we present a constructive proof that there exist complete pools for qubit-ADAPT containing only $2n-2$ operators, which we refer to as minimal complete pools. We first prove that the set of operators $\{V_i\}_{n=3}=\{iZ_3Z_2Y_1,iZ_3Y_2,iY_3,iY_2\}$ forms a complete pool for $n=3$ qubits, i.e., using operators of the form $\prod_k e^{\theta_kV_k}$, we can rotate any real state into any other real state. We then prove that the recursively defined pool, $\{V_i\}_n=\{Z_n\{V_i\}_{n-1},iY_n,iY_{n-1}\}$, is a complete pool for $n$ qubits, such that $\prod_k e^{\theta_kV_k}\ket{\psi}=\ket{\varphi}$ for any two real states $\ket{\psi}$ and $\ket{\varphi}$ in the Hilbert space.

\subsection{Complete pool for 3 qubits}

Here, we prove that $\{V_i\}_{n=3}{=}\{iZ_3Z_2Y_1,iZ_3Y_2,iY_3,iY_2\}$ is a minimal complete pool for three qubits. We will do this by showing that we can map any state $\ket{\psi}$ to $\ket{000}$ using only operators of the form $\prod_k e^{\theta_kV_k}$.

We begin by decomposing an arbitrary real state $\ket{\psi}$ as follows:
\be
\ket{\psi}=\ket{0}_3\ket{\psi_0}_{21}+\ket{1}_3\ket{\psi_1}_{21}.
\ee
The subscripts outside the kets indicate which qubit(s) the ket belongs to. In general, the two-qubit states $\ket{\psi_0}$ and $\ket{\psi_1}$ do not have the same norm: $\braket{\psi_0}{\psi_0}\ne\braket{\psi_1}{\psi_1}$. However, we can always perform a rotation $e^{i\theta Y_3}$ on qubit 3 (the left-most qubit) to transform $\ket{\psi}$ to the form
\be
|\psi'\rangle=e^{i\theta Y_3}\ket{\psi}=\ket{0}_3|\psi'_0\rangle_{21}+\ket{1}_3|\psi'_1\rangle_{21},\label{eq:psitilde}
\ee
where
\begin{align}
|\psi'_0\rangle&=\cos\theta\ket{\psi_0}+\sin\theta\ket{\psi_1},\\
|\psi'_1\rangle&=\cos\theta\ket{\psi_1}-\sin\theta\ket{\psi_0}.
\end{align}
The norms of these states are
\begin{align}
\langle\psi'_0|\psi'_0\rangle&=\cos^2\theta\braket{\psi_0}{\psi_0}+\sin^2\theta\braket{\psi_1}{\psi_1}\nonumber\\&+\sin(2\theta)\braket{\psi_0}{\psi_1},\\
\langle\psi'_1|\psi'_1\rangle&=\cos^2\theta\braket{\psi_1}{\psi_1}+\sin^2\theta\braket{\psi_0}{\psi_0}\nonumber\\&-\sin(2\theta)\braket{\psi_0}{\psi_1}.
\end{align}
We can make these two norms equal by choosing $\theta$ such that
\be
\tan(2\theta)=\frac{\braket{\psi_1}{\psi_1}-\braket{\psi_0}{\psi_0}}{2\braket{\psi_0}{\psi_1}}.
\ee
Note that this procedure works for arbitrary states $\ket{\psi_0}$ and $\ket{\psi_1}$, not just for two-qubit states.

The heart of the proof amounts to showing that a state of the form in Eq.~\eqref{eq:psitilde}, where $\langle\psi'_0|\psi'_0\rangle=\langle\psi'_1|\psi'_1\rangle=1/2$, can be transformed into a form where qubit 3 has been factored out:
\be
|\psi'\rangle\to(\ket{0}+\ket{1})_3\ket{\chi}_{21},\label{eq:3qubitmapping}
\ee
where $\ket{\chi}$ is some two-qubit state. We will do this using the (slightly reduced) set of pool operators $\{V^{\rm red}_i\}_3=\{iZ_3Z_2Y_1,iZ_3Y_2,iY_2\}$. These are the same as $\{V_i\}_3$, except that $iY_3$ is not included. In the next subsection, we generalize Eq.~\eqref{eq:3qubitmapping} to $n$ qubits and then use it to prove that $\{V_i\}_n$ is a complete pool.

To prove Eq.~\eqref{eq:3qubitmapping}, let's begin by expanding $|\psi'\rangle$ as follows:
\be
|\psi'\rangle=\ket{00}|\psi'_{0a}\rangle+\ket{01}|\psi'_{0b}\rangle+\ket{10}|\psi'_{1a}\rangle+\ket{11}|\psi'_{1b}\rangle.\label{eq:psitilde2}
\ee
From this point onward, we suppress the subscripts on the kets for notational simplicity. Similarly to how we made the states $|\psi'_0\rangle$ and $|\psi'_1\rangle$ have the same norm, here we can apply the conditional operations $e^{i\theta_0\frac{1+ Z_3}{2}Y_2}$ and $e^{i\theta_1\frac{1- Z_3}{2}Y_2}$ to make
\be
\langle\psi'_{0a}|\psi'_{0a}\rangle=\langle\psi'_{0b}|\psi'_{0b}\rangle=\langle\psi'_{1a}|\psi'_{1a}\rangle=\langle\psi'_{1b}|\psi'_{1b}\rangle=1/4.
\ee
We assume that this has already been done in Eq.~\eqref{eq:psitilde2}, and that $|\psi'\rangle$ has been redefined accordingly.

Next, we apply the conditional rotation $e^{i\phi_1Z_3Z_2Y_1}$ on qubit 1 to bring the state to
\be
e^{i\phi_1Z_3Z_2Y_1}|\psi'\rangle=\left(\ket{00}+\ket{01}\right)\ket{\chi_0}+\ket{10}|\psi''_{1a}\rangle+\ket{11}|\psi''_{1b}\rangle.\label{eq:conditionalrot}
\ee
Here, we used the fact that $|\psi'_{0a}\rangle$ and $|\psi'_{0b}\rangle$ are real single-qubit states, and thus can be viewed as 2D vectors lying in the same plane. The conditional rotation $e^{i\phi_1Z_3Z_2Y_1}$ rotates these two vectors in opposite directions because one is conditioned on the state $\ket{00}$ and the other on $\ket{01}$, and eventually the two vectors coincide at $\ket{\chi_0}$ for a particular value of the rotation angle $\phi_1$. This operation also affects the other terms in Eq.~\eqref{eq:conditionalrot} (as indicated with the extra primes), but their form remains the same. This conditional operation will be used repeatedly in what follows. 

The next step is to apply a conditional rotation on qubit 2 to bring it to the state $\ket{0}$ in the first term:
\begin{align}
e^{i\phi_2\frac{1+Z_3}{2}Y_2}e^{i\phi_1Z_3Z_2Y_1}|\psi'\rangle&=\sqrt{2}\ket{00}\ket{\chi_0}+\ket{10}|\psi''_{1a}\rangle\nonumber\\&+\ket{11}|\psi''_{1b}\rangle.
\end{align}
We can then apply another conditional operation on qubit 1 to rotate the states $|\psi''_{1a}\rangle$ and $|\psi''_{1b}\rangle$ into each other:
\begin{align}
&e^{i\phi_3Z_3Z_2Y_1}e^{i\phi_2\frac{1+Z_3}{2}Y_2}e^{i\phi_1Z_3Z_2Y_1}|\psi'\rangle\nonumber\\&=\sqrt{2}\ket{00}\ket{\chi'_0}+\left(\ket{10}+\ket{11}\right)\ket{\chi_1}.
\end{align}
This operation brings the two states to some state $\ket{\chi_1}$. It also changes $\ket{\chi_0}$ to $\ket{\chi'_0}$. Another conditional rotation on qubit 2 simplifies the second term:
\begin{align}
&e^{i\phi_4\frac{1-Z_3}{2}Y_2}e^{i\phi_3Z_3Z_2Y_1}e^{i\phi_2\frac{1+Z_3}{2}Y_2}e^{i\phi_1Z_3Z_2Y_1}|\psi'\rangle\nonumber\\&=\sqrt{2}\ket{00}\ket{\chi'_0}+\sqrt{2}\ket{10}\ket{\chi_1}.
\end{align}
We need one more conditional operation on qubit 1 to bring this to the desired form:
\begin{align}
&e^{i\phi_5Z_3Z_2Y_1}e^{i\phi_4\frac{1-Z_3}{2}Y_2}e^{i\phi_3Z_3Z_2Y_1}e^{i\phi_2\frac{1+Z_3}{2}Y_2}e^{i\phi_1Z_3Z_2Y_1}|\psi'\rangle\nonumber\\&=\sqrt{2}\left(\ket{00}+\ket{10}\right)\ket{\chi}=\left(\ket{0}+\ket{1}\right)\ket{0}\ket{\chi}\sqrt{2}.\label{eq:15}
\end{align}
We have thus proven Eq.~\eqref{eq:3qubitmapping}. For reasons that will become clear in the next subsection, it is important to note that we did this using only the reduced pool $\{V^{\rm red}_i\}_3$. 

All that remains is to show that Eq.~\eqref{eq:15} can be mapped to the state $\ket{000}$. This can be done using a $Y_3$ rotation followed by a conditional operation on qubit 1:
\begin{align}
&e^{i\phi_6Z_3Z_2Y_1}e^{i\frac{\pi}{4}Y_3}e^{i\phi_5Z_3Z_2Y_1}e^{i\phi_4\frac{1-Z_3}{2}Y_2}e^{i\phi_3Z_3Z_2Y_1}e^{i\phi_2\frac{1+Z_3}{2}Y_2}\nonumber\\&\times e^{i\phi_1Z_3Z_2Y_1}|\psi'\rangle=\ket{000}.
\end{align}
Since we can map any state to $\ket{000}$, it follows that we can map any 3-qubit state to any other 3-qubit state using only the pool operators.
We have thus shown that $\{V_i\}_3=\{iZ_3Z_2Y_1,iZ_3Y_2,iY_3,iY_2\}$ is a minimal complete pool for three qubits. 

\subsection{Proof of the complete pool for $n$ qubits}

In the previous subsection, we proved that the pool $\{V_i\}_3$ is a complete pool for three qubits. A key piece of this proof is the statement that we can use the reduced pool $\{V^{\rm red}_i\}_3=\{iZ_3Z_2Y_1,iZ_3Y_2,iY_2\}$ to factor out the 3rd qubit Eq.~\eqref{eq:3qubitmapping}:
\be
\ket{0}|\psi'_0\rangle+\ket{1}|\psi'_1\rangle\to\left(\ket{0}+\ket{1}\right)\ket{\chi},
\ee
where we assume $\langle\psi'_0|\psi'_0\rangle=\langle\psi'_1|\psi'_1\rangle$, and $\ket{\chi}$ is some 2-qubit state. We will prove that a similar statement holds for $n$ qubits:

~

\textbf{Theorem:} We can factor out the $n$th qubit from a generic $n$-qubit state:
\be
\ket{0}|\psi'_0\rangle+\ket{1}|\psi'_1\rangle\to\left(\ket{0}+\ket{1}\right)\ket{\chi},
\ee
where $\langle\psi'_0|\psi'_0\rangle=\langle\psi'_1|\psi'_1\rangle$, and $\ket{\chi}$ is some $(n-1)$-qubit state, using only the reduced $n$-qubit pool:
\begin{align}
\{V^{\rm red}_i\}_n&=\{Z_n\{V_i\}_{n-1},iY_{n-1}\}\nonumber\\&=\{Z_n\{V^{\rm red}_i\}_{n-1},iZ_nY_{n-1},iY_{n-1}\}.
\end{align}
This pool is defined recursively starting from $\{V^{\rm red}_i\}_3$, and it differs from the full pool by only one operator $(iY_n)$:
\be
\{V_i\}_n=\{Z_n\{V_i\}_{n-1},iY_n,iY_{n-1}\}.
\ee

We will use induction to prove this theorem. We will assume it holds for $n$ qubits and then show that it also holds for $n+1$ qubits. This proof will closely follow what we did for three qubits above. Thus, we start with the $(n+1)$-qubit state:
\be
\ket{0}_{n+1}|\psi'_0\rangle+\ket{1}_{n+1}|\psi'_1\rangle,
\ee
where $|\psi'_0\rangle$ and $|\psi'_1\rangle$ are $n$-qubit states that have the same norm. We want to factor out the left-most $(n+1)$th qubit using only the operators $\{V^{\rm red}_i\}_{n+1}=\{Z_{n+1}\{V^{\rm red}_i\}_{n},iZ_{n+1}Y_n,iY_n\}$. We decompose the state further:
\be
\ket{00}|\psi'_{0a}\rangle+\ket{01}|\psi'_{0b}\rangle+\ket{10}|\psi'_{1a}\rangle+\ket{11}|\psi'_{1b}\rangle.
\ee
As before, we can assume that all four states $|\psi'_{0a}\rangle$, $|\psi'_{0b}\rangle$, $|\psi'_{1a}\rangle$, $|\psi'_{1b}\rangle$, have the same norm of $1/4$, since this can be achieved by applying two conditional operations $e^{i\theta_0\frac{1+Z_{n+1}}{2}Y_n}$ and $e^{i\theta_1\frac{1-Z_{n+1}}{2}Y_n}$.

We next apply a conditional operation built from the operators $e^{\phi_1Z_{n+1}\{V^{\rm red}_i\}_n}$ to map the state to
\be
\left(\ket{00}+\ket{01}\right)\ket{\chi_{0}}+\ket{10}|\psi''_{1a}\rangle+\ket{11}|\psi''_{1b}\rangle.
\ee
We used the induction hypothesis for $n$ qubits to obtain this result. Since the left-most qubit is in the state $\ket{0}$ in the first two terms, the effect of $e^{\phi_1Z_{n+1}\{V^{\rm red}_i\}_n}$ is the same as $e^{\phi_1\{V^{\rm red}_i\}_n}$ on these terms, and the hypothesis for $n$ qubits then allows us to bring these two terms into a factorized form as above. The last two terms also evolve under this operation, but their general form remains the same since the $n+1$ qubit remains in the state $\ket{1}$ under this operation. We follow this with a conditional operation $e^{i\phi_2\frac{1+Z_{n+1}}{2}Y_n}$ on the $n$th qubit to rotate it to $\ket{0}$ in the first term:
\be
\sqrt{2}\ket{00}\ket{\chi_{0}}+\ket{10}|\psi''_{1a}\rangle+\ket{11}|\psi''_{1b}\rangle.
\ee
Another application of $e^{\phi_3Z_{n+1}\{V^{\rm red}_i\}_n}$ can be used to similarly factorize the last two terms:
\be
\sqrt{2}\ket{00}\ket{\chi'_{0}}+\left(\ket{10}+\ket{11}\right)\ket{\chi_1}.
\ee
Note that the form of the first term remains the same since this conditional operation does not rotate the $(n+1)$th or $n$th qubits. This is because all the operators in $\{V^{\rm red}_i\}_n$ have either a $Z$ or a $I$ on the $n$th qubit.
Next apply $e^{i\phi_4\frac{1-Z_{n+1}}{2}Y_n}$ to rotate the $n$th qubit in the second term to $\ket{0}$:
\be
\sqrt{2}\ket{00}\ket{\chi'_{0}}+\sqrt{2}\ket{10}\ket{\chi_1}.\label{eq:almostfactorized}
\ee
Additional applications of $e^{\phi Z_{n+1}\{V^{\rm red}_i\}_n}$ bring this to
\be
\sqrt{2}\left(\ket{0}+\ket{1}\right)\ket{0}\ket{\chi},\label{eq:factorized}
\ee
as desired. In order to make this final step, we need to use a slight extension of the theorem in which we factor out the $(n+1)$th qubit instead of the $n$th qubit. Do we have all the necessary operators to make this extension work? Yes. To see this, recall that in order to factor out the $n$th qubit, we need $\{V^{\rm red}_i\}_n=\{Z_n\{V^{\rm red}_i\}_{n-1},iZ_nY_{n-1},iY_{n-1}\}$. However, here we have
\begin{align}
&Z_{n+1}\{V^{\rm red}_i\}_n=\nonumber\\&\{Z_{n+1}Z_n\{V^{\rm red}_i\}_{n-1},iZ_{n+1}Z_nY_{n-1},iZ_{n+1}Y_{n-1}\}.\label{eq:nplus1ops}
\end{align}
The operators $Z_{n+1}Z_n\{V^{\rm red}_i\}_{n-1}$ and $iZ_{n+1}Z_nY_{n-1}$ produce exactly the same conditional rotations on the state in Eq.~\eqref{eq:almostfactorized} as $Z_n\{V^{\rm red}_i\}_{n-1}$ and $iZ_nY_{n-1}$ produce on $\sqrt{2}\ket{0}\ket{\chi'_{0}}+\sqrt{2}\ket{1}\ket{\chi_1}$. However, the theorem also requires unconditional $Y_{n-1}$ rotations on $\ket{\chi'_0}$ and $\ket{\chi_1}$. These can be implemented using the composite operator
\be
e^{i\frac{\pi}{4}\frac{1-Z_{n+1}}{2}Y_{n}}e^{i\phi Z_{n+1}Z_nY_{n-1}}e^{i\frac{\pi}{4}\frac{1-Z_{n+1}}{2}Y_{n}},\label{eq:compositeYn-1}
\ee
which first rotates \eqref{eq:almostfactorized} into a state in which the $(n+1)$th and $n$th qubit states in the two terms have the same parity before performing the $Z_{n+1}Z_nY_{n-1}$ rotation, so that $\ket{\chi'_{0}}$ and $\ket{\chi_1}$ undergo the same rotation. The $(n+1)$th and $n$th qubits are then restored to their original states by the final rotation in Eq.~\eqref{eq:compositeYn-1}. Given that we have already shown that this procedure works for $n=3$ qubits, this completes the proof of the theorem.

With the theorem proven, it is easy to prove that $\{V_i\}_{n+1}$ is a complete pool. We already know that a rotation generated by $Y_{n+1}$ is enough to bring an arbitrary $(n+1)$-qubit state to the form needed to apply the theorem. As we just saw, we can then use $Z_{n+1}\{V^{\rm red}_i\}_{n}$, $iZ_{n+1}Y_{n}$, and $iY_{n}$ to factor out the $(n+1)$th qubit and arrive at Eq.~\eqref{eq:factorized}. If we then apply another $Y_{n+1}$ rotation to bring the $(n+1)$th qubit to $\ket{0}$, we can perform the conditional operation $e^{i\phi_6Z_{n+1}\{V^{\rm red}_i\}_n}$ to rotate $\sqrt{2}\ket{\chi}$ to $\ket{0}^{\otimes n-1}$. Thus, we can rotate an arbitrary $(n+1)$-qubit state to $\ket{0}^{\otimes n+1}$ using only the generators $\{V_i\}_{n+1}=\{Z_{n+1}\{V^{\rm red}_i\}_n,iZ_{n+1}Y_n,iY_{n+1},iY_n\}$. This in turn implies that we can rotate any $(n+1)$-qubit state to any other $(n+1)$-qubit state using the same set of operators. Therefore, $\{V_i\}_{n+1}$ is a complete pool, and it contains $2n-2$ operators.\\

\section{Mapping between minimal complete pools}\label{app:mappools}

In this section, we show that the two examples of minimal complete pools discussed in the main text can be mapped into one another. The first example is the pool $\{V_i\}$ that is constructed iteratively in Appendix~\ref{app:proof}. The second example is the pool comprised of operators that only act on adjacent qubits in a linear array, $\{G_i\}$. To facilitate the following analysis, we order the elements in these two pools as follows (where we have ignored the factors of $i$ in the definition of the pool operators for simplicity):
\begin{widetext}
\begin{align}
    &V_1=ZZ\ldots ZY,\quad V_2=ZZ\ldots ZYI,\quad V_3=ZZ\ldots ZYII,\quad\ldots,\quad V_{n-1}=ZYII\ldots I,\quad V_n=YII\ldots I,\nonumber\\
    &V_{n+1}=ZZ\ldots ZIYI,\quad V_{n+2}=ZZ\ldots ZIYII,\quad\ldots,\quad V_{2n-3}=ZIYII\ldots I,\quad V_{2n-2}=IYII\ldots I.
\end{align}
\begin{align}
    &G_1=ZYII\ldots I,\quad G_2=IZYII\ldots I,\quad G_3=IIZYII\ldots I,\quad\ldots,\quad G_{n-2}=II\ldots IZYI,\quad G_{n-1}=II\ldots IZY,\nonumber\\
    &G_{n}=YII\ldots I,\quad G_{n+1}=IYII\ldots I,\quad G_{n+2}=IIYII\ldots I,\quad\ldots,\quad G_{2n-3}=II\ldots IYII,\quad G_{2n-2}=II\ldots IYI.
\end{align}
\end{widetext}
We will show that the $V_i$ can be obtained from commutators of the $G_i$. This means that the two pools share the same operator basis, and so completeness of the one pool implies completeness of the other. Since we have already proven that the $V_i$ are complete in Appendix~\ref{app:proof}, this demonstrates that the $G_i$ also form a minimal complete pool for any number $n$ of qubits.

Recall that two Pauli strings do not commute if they differ by an odd number of Pauli operators, ignoring any qubits with identity operators in at least one of the two strings. In this case, the result of the commutator is proportional to the product of the two Pauli strings. Now notice that we can write the $V_i$ as products of the $G_i$:
\begin{align}
    V_{n-k}&\sim G_1\prod_{j=2}^kG_jG_{j+n-1},\quad k=2,\ldots,n-1,\nonumber\\
    V_{n-1}&=G_{1},\nonumber\\
    V_n&=G_{n},\nonumber\\
    V_{2n-k}&\sim G_{k-1}G_{k}V_{n-k},\quad k=3,\ldots,n-1,\nonumber\\
    V_{2n-2}&=G_{n+1}.
\end{align}
The operators in the product in the first line should be ordered such that the operators with the smallest values of $j$ are on the left and the ones with the largest values are on the right. Notice that no two adjacent operators in these products commute. Therefore, we can rewrite them as nested commutators of the $G_i$. This proves that the $V_i$ and $G_i$ produce the same operator basis when all possible commutators of each pool are computed. We already showed in Appendix~\ref{app:proof} that the $V_i$ are complete. Therefore, the $G_i$ also constitute a minimal complete pool. This pool is particularly useful for quantum processors containing a linear array of qubits with nearest-neighbor coupling only. Interestingly, the entangling operators in the $G_i$ pool are similar to the cross-resonance interaction in fixed-frequency superconducting qubits, such as those in the IBM quantum processors \cite{Rigetti2010PRB, Chow2011PRL, Sheldon2016PRA}. This similarity could be exploited to implement qubit-ADAPT with native hardware operations.

%


\end{document}